\documentclass[aps,twocolumn,amsmath,superscriptaddress,prd,floatfix]{revtex4-2}
\usepackage{graphicx,color,dcolumn,booktabs,bm}
\usepackage{longtable}
\usepackage{txfonts}
\usepackage{overpic}
\usepackage{epsfig}
\usepackage{amssymb}
\usepackage{epstopdf}
\usepackage{indentfirst}
\usepackage{feynmf}
\usepackage{slashed}
\usepackage{cases}
\usepackage{color}
\usepackage{multirow}
\usepackage{graphicx,color,dcolumn,booktabs,bm}
\usepackage{cases}
\usepackage{array}

\graphicspath{{Figures/}} 

\usepackage[colorlinks,citecolor=blue,anchorcolor=red,menucolor=red,linkcolor=red,filecolor=red,runcolor=red,urlcolor=blue,frenchlinks=red]{hyperref}

\begin{document}

\title{$\eta$ and $\eta'$ mesons from $N_f = 2+1$ lattice QCD at the physical point using topological charge operators}

\author{Yue Su} %
\affiliation{Department of Physics, Hunan Normal University,  Changsha 410081, China }
\affiliation{Center for Joint Quantum Studies and Department of Physics, School of Science, Tianjin University, Tianjin 300350, China}
\author{Nan Wang} %
\affiliation{State Key Laboratory of Nuclear Physics and Technology, Institute of Quantum Matter, South China Normal University, Guangzhou 510006, China}
\affiliation{Key Laboratory of Atomic and Subatomic Structure and Quantum Control~(MOE), Guangdong-Hong Kong Joint Laboratory of Quantum Matter, Guangzhou 510006, China}
\affiliation{Guangdong Basic Research Center of Excellence for Structure and Fundamental Interactions of Matter, Guangdong Provincial Key Laboratory of Nuclear Science, Guangzhou 510006, China}
\author{Long-cheng Gui} %
\email{guilongcheng@hunnu.edu.cn}
\affiliation{Department of Physics, Hunan Normal University,  Changsha 410081, China }
\affiliation{Hunan Research Center of the Basic Discipline for Quantum Effects and Quantum Technologies, Hunan Normal University, Changsha 410081, China}
\affiliation{Synergetic Innovation
Center for Quantum Effects and Applications (SICQEA), Changsha 410081, China}
\affiliation{Key Laboratory of Low-Dimensional Quantum Structures and Quantum Control of Ministry of Education, Changsha 410081, China\\~\\
\includegraphics[scale=0.1]{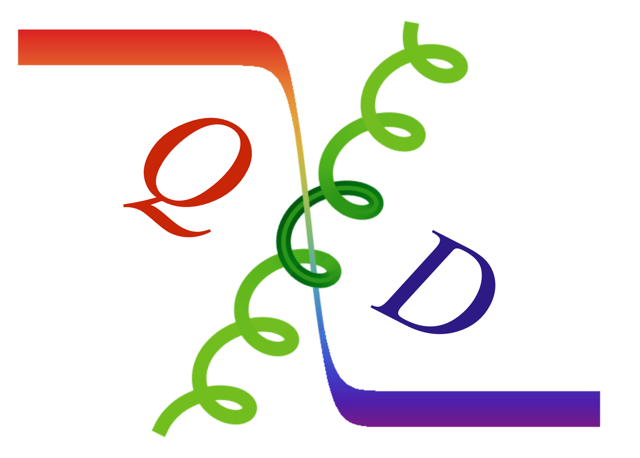}\\
{\rm($\chi$QCD Collaboration)}}
\author{Jun Hua} %
\author{Jian Liang} %
\email{jianliang@scnu.edu.cn}
\author{Jun Shi} %
\email{jun.shi@scnu.edu.cn}
\affiliation{State Key Laboratory of Nuclear Physics and Technology, Institute of Quantum Matter, South China Normal University, Guangzhou 510006, China}
\affiliation{Guangdong Basic Research Center of Excellence for Structure and Fundamental Interactions of Matter, Guangdong Provincial Key Laboratory of Nuclear Science, Guangzhou 510006, China}

\begin{abstract}
By fitting the two-point correlation functions of topological charge density operators calculated on two $2+1$-flavor gauge ensembles with physical pion mass, we determine both the $\eta$ and $\eta'$ masses and also the mixing angle to be $m_\eta = 0.505(72)(75)$ GeV, $m_{\eta'}=0.952(47)(40)$ GeV, and $\theta_1 = -8.9(2.1)(1.8)^\circ$, respectively, where the first error is the statistical uncertainty and the second one is the systematic uncertainty.
This is the first extraction of both $\eta/\eta'$ masses and the mixing angle $\theta_1$ using topological charge operators.
Compared with previous studies using quark bilinear operators, the error of the $\eta$ mass is relatively large, but the mixing angle has comparable precision.
This demonstrates that the topological charge operators are well suited to study the $\eta$ and $\eta'$ mesons.
\end{abstract}
\maketitle

\section{Introduction}{\label{introduction}}
Understanding the masses of $\eta$ and $\eta'$ mesons as well as the mixing between the original flavor singlet state $\eta_1$ and the neutral octet state $\eta_8$ is an important topic in low-energy strong-interaction physics. 
In one respect, the spontaneous breaking of the chiral symmetry results in the mass of the Goldstone bosons. 
Also, through the partially conserved axial current (PCAC) relation, the pseudoscalar current is modified by the non-trivial topological structure of the gauge fields, and the mass of the flavor-singlet pseudoscalar particle is significantly heavier than that of the other pseudoscalar particles due to the $\rm U_A(1)$ anomaly. 
In another respect, the mixing angle between the flavor singlet and octet states is a direct measurement of the breaking of $\rm SU(3)$ flavor symmetry~\cite{Weinberg:1975ui,Belavin:1975fg,Witten:1979vv}.

Due to the flavor-singlet components, lattice calculations of $\eta$ and $\eta'$ mesons using quark bilinear interpolating field operators involve disconnected insertions, which require high statistics. Though challenging, many efforts with this method have been made in the community on the $\eta$, $\eta'$ mesons, from quenched approximations~\cite{Kuramashi:1994aj, Venkataraman:1997xi} to dynamical fermions with $N_f = 2$~\cite{McNeile:2000hf,TXL:2000mhy,Bali:2001gk,CP-PACS:2002exu,Hashimoto:2008xg,Jansen:2008wv,Jiang:2022ffl}, $N_f = 2+1$~\cite{Christ:2010dd,Dudek:2011tt,Gregory:2011sg,Bali:2014pva,Cichy:2015jra,Bali:2021qem,CSSMQCDSFUKQCD:2021rvs,Verplanke:2024msi}, and $N_f = 2+1+1$~\cite{Ottnad:2012fv,Michael:2013gka,Ottnad:2017bjt,Ottnad:2025zxq}. Moreover, the study~\cite{Christ:2010dd} reports the first realistic calculation of the octet-singlet mixing angle and confirms that the large mass of $\eta'$ arises from the combined
effects of the axial anomaly and the gauge field topology in QCD. Recent works~\cite{Bali:2021qem,Ottnad:2025zxq} provide more detailed calculations and analysis on this topic with gauge ensembles of different parameters, including ones with the physical pion mass.

The ingenious calculation of $\eta'$ mass using operators of the topological charge density was carried out in 2015 on a 2+1-flavor lattice at one single lattice spacing and with an unphysical pion mass~\cite{Fukaya:2015ara}.
This approach
avoids the difficulties of handling disconnected insertions.
A follow-up work~\cite{Dimopoulos:2018xkm} performs similar calculations of the $\eta'$ meson mass on several $N_f = 2$ flavor gauge ensembles with one at the physical pion mass.

For now, there still lacks a comprehensive lattice calculation of both $\eta$ and $\eta'$ masses and also the mixing angle using the interpolating operators of topological charge density. 
Such a study provides insights into the nature of $\eta$ and $\eta'$ mesons, and also the physics of axial anomaly and gauge field topology. 
The results can be compared with those obtained using quark bilinear operators.
In this paper, we report our calculations of the masses of the $\eta$ and $\eta'$ mesons as well as the singlet-octet mixing angle $\theta_1$ using topological charge density operators on two $2+1$-flavor gauge ensembles both at the physical point with different lattice spacings. 
A simple continuum limit is taken and the systematic uncertainties in the data analysis are carefully considered.

This paper is organized as follows: In Sec.~\ref{sec2}, we describe our lattice setup and discuss the two-point correlation functions of the topological charge density operators. Sec.~\ref{sec3} presents the details of data analysis on the $\eta$ and $\eta'$ masses and the mixing angle. Some discussion and a summary are provided in Sec.~\ref{sec4}.

\section{NUMERICAL DETAILS}\label{sec2} 
We employ two $2+1$-flavor gauge ensembles of domain wall fermions generated by the RBC/UKQCD collaboration~\cite{RBC:2014ntl}.
Both of them are at the physical pion mass point.
We have two different lattice spacings, and the discretization effects of our results are carefully estimated.
Detailed parameters of the gauge ensembles are collected in Table~\ref{tab:1}.

\begin{table}[ht]
\centering
\begin{tabular}{lccccc}
\hline \hline Symbol & $L^3 \times T$ & $a$ $(\mathrm{fm})$ & $m_K$ (MeV) & $m_\pi$ (MeV)& $N_{\rm cfg}$ \\
\hline $48 \mathrm{I}$ & $48^3 \times 96$ & $0.1141(2)$ & 499 & 139 & 356 \\
$64 \mathrm{I}$ & $64^3 \times 128$ & $0.0837(2)$ & 508 & 139 & 330 \\
\hline\hline
\end{tabular}
\caption{Parameters of the gauge ensembles including the numbers of the configurations $N_{\rm cfg}$ used in this study.}
\label{tab:1}
\end{table}

As stated in the introduction, in this work, we extract the masses of the $\eta$ and $\eta'$ mesons and the corresponding mixing parameter from calculating the correlation functions of topological charge density operators. 
The topological charge is actually the scalar product of the color electric and magnetic fields and carries a quantum number the same as $\eta$ and $\eta'$, i.e., $0^+0^{-+}$.
We use the Symanzik improved version of the topological charge density to
reduce the discretization errors~\cite{Alexandrou:2015yba},
\begin{align}\label{eq:topo-density}
q(x)=c_0 q^{\text {clov}}(x)+c_1 q^{\text {rect }}(x),
\end{align}
where $c_0=5 / 3$ and ${c}_1=-1 / 12$ are the tree-level Symanzik coefficients.
$q^{\text {clov}}(x)$ and $q^{\text {rect}}(x)$ are defined from the corresponding
field tensor $F_{\mu \nu}^{\text{clov~(rect)}}(x)$ as
\begin{align}
q(x)^{\text{clov~(rect)}}=\frac{1}{32 \pi^2} \epsilon_{\mu \nu \rho \sigma} \operatorname{Tr}\left\{F_{\mu \nu}^{\text{clov~(rect)}}(x)F_{\rho \sigma}^{\text{clov~(rect)}}(x)\right\},
\end{align}
where $F_{\mu\nu}^{\text{clov}}(x)$ represents the ordinary clover leaf construction while $F_{\mu\nu}^{\text{rect}}(x)$ is from the sum of the horizontally and vertically oriented rectangular Wilson loops.

\begin{figure}[htbp]
\centering
\includegraphics[width=0.35\textwidth]{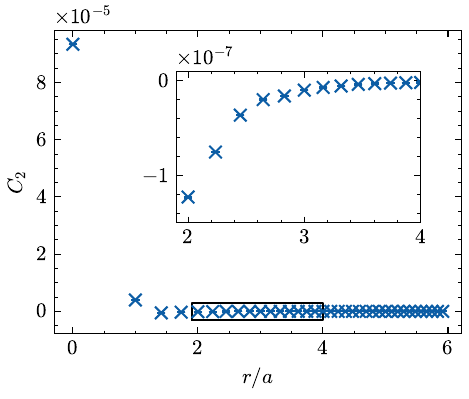}
\centering
\caption{The two-point correlation function of the topological charge density operator $C_2(r)$ as a function of $r$ on the 48I ensemble. The zoomed in figure shows
the negative correlation at large $r$.}
\label{corr48}
\end{figure}

The Euclidean correlation functions of topological charge density operators are expressed as
\begin{align}
    C_{2}(r) = \frac{1}{VN_r}\sum_{y\in\{|y - x|=r|\}}\sum_x ~ \langle q(y)~q(x) \rangle,
\end{align}
where $x$ and $y$ are spacetime coordinates, $r=|y - x|$ is a 4-D distance,
$V$ is the lattice volume, and $N_r$ is the number of $y$ that gives $r$ for a given $x$. The correlator $C_{2}$ can be calculated effectively 
by means of convolution theorem and fast Fourier transform. 
The reason we calculate the 4-D correlation of the topological charge operators rather than the common temporal correlation lies in the behavior of the correlation functions.
As shown in Fig.~\ref{corr48}, the two-point correlation function has a positive kernel at small $r$ and changes its sign with increasing $r$. 
Actually, in the continuum, the reflection positivity and the fact that the topological charge is reflection-odd ensure that the topological charge correlation functions are negative at non-zero distance, and the positive contact contribution is a delta function at $|x-y|=0$. However, the topological charges are not locally defined on the lattice, the positive contact contribution is smeared to cover several lattice spacings~\cite{Horvath:2002yn, Horvath:2005cv}. 
Therefore, in order to reliably extract the physical states, one needs to skip the first several lattice spacings and use only the data points of the negative tail. 
In such a situation, calculating the 4-D distance provides more usable data points, which is also a common choice in previous studies~\cite{Fukaya:2015ara, Dimopoulos:2018xkm}.

Since we are considering 4-D distances, we have many data points at very close $r$ when $r$ is large. 
These data points can be strongly correlated and result in difficulties in the numerical analysis. 
Therefore, we perform a binning procedure: we average the data points in each interval of $\delta_r=0.5a$ in $r$ as a binned data point. 
The distance after binning is denoted as $\hat{r}=(0.25+i\times 0.5)a$ for the $i$-th interval.

\begin{figure}[htbp]
    \centering    
    \includegraphics[width=0.35\textwidth]{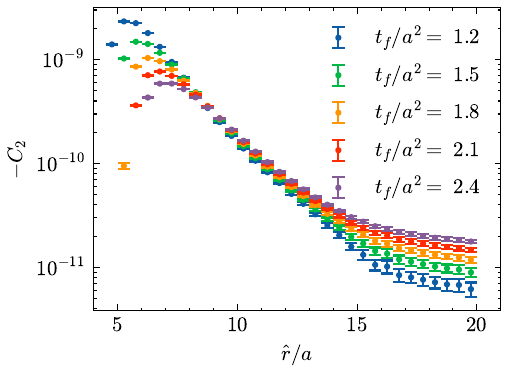}
    \caption{The correlation functions of topological charge density operators on the 64I ensembles.
    Additional minus sign is included and the data points with different flow times are shifted vertically for clarity.}
    \label{corr64_minus}
\end{figure}

To improve the statistical signal-to-noise ratio, we use Wilson flow~\cite{Luscher:2010iy, Luscher:2011bx} to smear our gauge fields. 
For a flow time $t_f$, this process creates effectively a smearing radius of $\sqrt{8t_f}$. The application of Wilson flow results in the suppression of the short-range fluctuations and statistical noises~\cite{Chowdhury:2014mra}.
Besides, we use jackknife resampling method to handle the
auto-correlation effects of the gauge configurations,
and to obtain reliably estimated statistical errors.
Fig.~\ref{corr64_minus} displays the topological charge density correlator ($-C(r)$ actually for better readability) on the 64I ensemble at different flow times $t_f/a^2$ = 1.2, 1.5, 1.8, 2.1 and 2.4. 
Since for large $\hat{r}$, the correlators of different flow times overlap with each other, we shift the data points vertically for clarity.
It is evident that, as the flow time $t_f$ increases, the signal-to-noise ratios at large $r$ get improved. 
However, in the meanwhile, the contact contribution is further smeared and one has fewer data points left to use. 
Thus, as a compromise, we find that $t_f/a^2$=2.4 is a good choice. In fact, the errors at $t_f/a^2$=2.1 and $t_f/a^2$=2.4 are quite close, and the numerical results extracted from different flow times are all consistent within errors. 
For the 48I lattice, the optimal flow time we find is $t_f/a^2$=1.6. 
The difference stems from their different lattice spacings. 

\begin{figure}[htbp]
    \centering
    \includegraphics[width=0.35\textwidth]{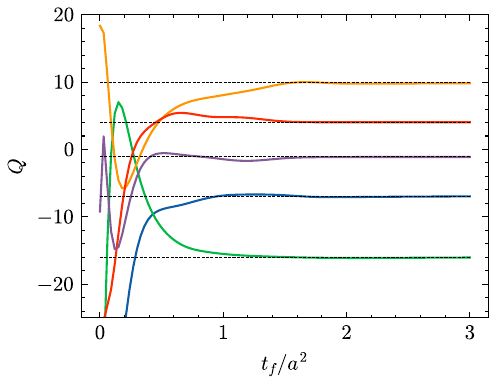}
    \caption{The topological charges of several different gauge configurations as a function of flow time on the 64I lattice. The horizontal dashed lines indicate the nearest integers.}
    \label{fig:topochg}
\end{figure}

Moreover, Wilson flow is essential to properly define the topological charge operator, which is important in the extraction of the mixing angle.
The topological charge of a gauge field configuration should be integer-valued in the continuum.
However, on the lattice, direct calculations of the topological charge from the $F\tilde F$ definition will not yield integer values due to the discretization effects.
Proper smearings that suppress the UV fluctuations and ``smooth" the gauge fields help obtain near-integer topological charges.
As shown in Fig.~\ref{fig:topochg}, when $t_f$ is large enough ($t_f/a^2\gtrsim1.6$ on this ensemble), the topological charges do saturate and approach near-integer values. 
Further smearings do not change the operators or correlators significantly. 
So the flow times we choose to use are also appropriate in this sense.

\section{Fittings and Results} \label{sec3}
According to the form of the scalar propagator in coordinate space with space-like separation, the correlation functions of topological charge density operators
can be expressed as
\begin{equation}
   C_2(\hat{r})=\sum_n A_n \frac{m_n}{4\pi^2\hat{r}}K_{1}(m_n \hat{r}), 
   \label{fit_f}
\end{equation}
where $K_{1}$ is the modified Bessel function of the second kind, and the summation runs over all the states with $0^+0^{-+}$ quantum number.
In the practical calculations, we use the asymptotic expansion of 
$K_{1}(z)=\sqrt{\frac{\pi}{2z}}e^{-z}\left(1+\frac{3}{8z}\right)$.
In the range of $z$ we consider, the difference is around or less than one percent,
so the precision is enough for this study.

\begin{figure}[htbp]
    \centering \includegraphics[width=0.35\textwidth]{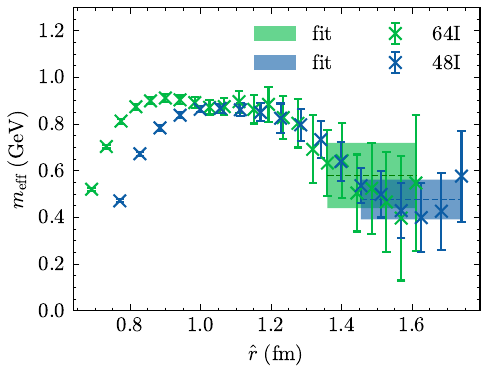}
    \caption{The effective masses and fitting results of the $\eta$ mass. The bands also indicate the fitting ranges.}
    \label{eff_all}
\end{figure}

To have qualitative information of the spectrum, we first extract the effective mass from the correlation functions. 
The effective mass at each $\hat{r}$ is obtained by solving the following equation numerically
\begin{equation}
    \frac{K_1(m_{\rm eff}(\hat{r}-\delta_r))}{K_1\left(m_{\rm eff}(\hat{r}+\delta_r)\right)}\frac{\hat{r}+\delta_r}{\hat{r}-\delta_r} = \frac{C_2(\hat{r}-\delta_r)}{C_2(\hat{r}+\delta_r)}.
    \label{ratio}
\end{equation}
The results are shown in Fig.~\ref{eff_all}.
One finds that, when $\hat{r}$ is larger than $\sim$ 1.4 fm, the effective masses of both lattices are close to the $\eta$ mass. 
And in the middle $\hat{r}$ region, the effective masses are close to the $\eta'$ mass.
For very small $\hat{r}$, unphysical contributions due to the positive kernel appear.
These observations demonstrate that the correlation functions indeed contain information from both $\eta$ and $\eta'$ states.
Moreover, they inspired an approach to extract the information:
first perform single-state fits in the large $\hat{r}$ region to determine the mass and spectral weight of the ground-state $\eta$ and then apply constrained two-state fits in the intermediate $\hat{r}$ region to fix the excited-state $\eta'$'s parameters. 
Details are described in the following subsections.

\subsection{fits for the $\eta$ state} 
\begin{figure}[htbp]
    \centering \includegraphics[width=0.48\textwidth]{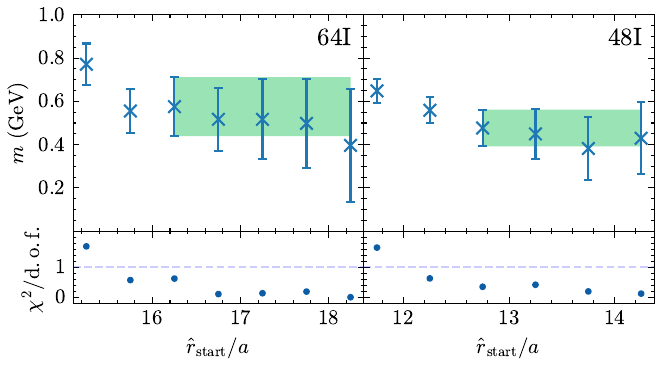}
    \caption{Left panel: single-state fitting results for the $\eta$ mass on the 64I ensemble as a function of the fitting starting point $\hat{r}_{\rm start}$. The corresponding $\chi^2/\text{d.o.f.}$ are also presented.
    The bands are to indicate the picked fitting and to show
    more clearly the difference of the fitting results between different $\hat{r}_{\rm start}$.
    Right panel: the same plot but for the 48I ensemble.
    }
    \label{eta_64}
\end{figure}

We use the single-state version of Eq.~\eqref{fit_f} to fit for the $\eta$ state. The fitting results as a function of different starting points $\hat{r}_{\rm start}$ together with the corresponding $\chi^2/\text{d.o.f.}$ are depicted in Fig.~\ref{eta_64}.
There are relatively strong correlations among the data points ($\sim$90\% for adjacent points and decreases to $\sim$50\% when the points are separated by two intermediate points). Those correlations are properly handled by using truncated SVD when inverting the covariance matrix. 
We follow the approach provided in the lsqfit package~\footnote{\noindent https://github.com/gplepage/lsqfit} to set the SVD cut, in which the basic idea is identifying poorly estimated eigenmodes of the correlation matrix~\cite{Dowdall:2019bea}. The endpoints of the fits are fixed to be $19.25a$ and $15.25a$ for the 64I and 48I ensembles, respectively, where the signal starts to be overwhelmed by noise. We identify that the results saturate at $\hat{r}_{\rm start}/a=16.25$ for the 64I lattice and $\hat{r}_{\rm start}/a=12.75$ for the 48I lattice (green bands in the figures). By checking the fitting results of the larger $\hat{r}_{\rm start}$, we confirm that the corresponding systematic uncertainties are subdominant to the statistical ones and thus are not included in the results.

The resulting $\eta$ mass is 0.58(14) GeV on the 64I lattice and 0.477(85) GeV on the 48I lattice. The results are plotted in Fig.~\ref{eff_all} also for sanity check.
The fitting results are well consistent with the effective mass data point for both the lattices.
Since these two values are consistent within errors, we conclude that, within the current statistical precision, no significant discretization errors are observed. 
This finding is consistent with our earlier results using chiral fermions, e.g., Ref.~\cite{Liang:2018pis}.
Therefore, we determine our final prediction by using a constant fit of the results from the two lattices. 
Furthermore, we take the difference between the constant-fit result and the value obtained on the finer lattice as the systematic uncertainty.
In the end, we get $m_\eta = 0.505(72)(75) $ GeV, which is consistent with the experimental value of 547.862(17) MeV~\cite{ParticleDataGroup:2024cfk}. More detailed comparisons are collected in Fig.~\ref{fig:compare_results}.

\subsection{fits for the $\eta'$ state} 

We use Eq.~\eqref{fit_f} with two states to fit for the $\eta'$ state
\begin{equation}\label{eq:2mass-fit}
    C_2(\hat{r}) =A_{\eta'} \frac{m_{\eta^{\prime}}}{4\pi^2\hat{r}}  K_{1}\left(m_{\eta^{\prime}} \hat{r}\right) 
    +A_{\eta}\frac{m_{\eta}}{4\pi^2\hat{r}} K_{1}\left(m_{\eta} \hat{r}\right).
\end{equation}
To make the fits more stable, the previously fitted results of $\eta$ mass are used as priors in these two-state fits. 
The spectral weight factor $A_{\eta}$ is free of prior since the $A_{\eta}$'s from the single-state fits are not as reliable as the masses.
$N.B.$: In Bayesian statistics, the prior should be independent of the evidence. Therefore, we use different data points in the $\eta'$ fitting.
Specifically, the end points of the $\eta'$ fittings are fixed to be $13.75a$ and $11.75a$   for the 64I and 48I ensembles respectively, which are actually the preceding points of the starting points in the fits of $\eta$.

\begin{figure}[htbp]
    \centering \includegraphics[width=0.48\textwidth]{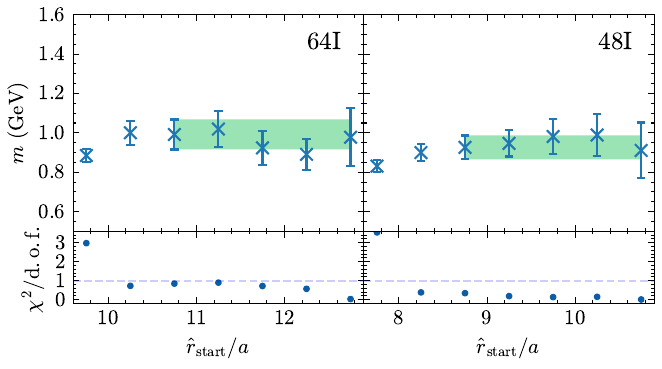}
    \caption{Same as Fig.~\ref{eta_64} but for the $\eta'$ sates.
    }
    \label{etap_64}
\end{figure}

Similarly, the fitting results as a function of different starting points $\hat{r}_{\rm start}$
together with the corresponding $\chi^2/\text{d.o.f.}$ are depicted in Fig.~\ref{etap_64}.
The results saturate at $\hat{r}_{\rm start}/a=10.75$ for the 64I lattice and $\hat{r}_{\rm start}/a=8.75$ for the 48I lattice (green bands in the figures).
Again, the $\chi^2/\text{d.o.f.}$ in each case is less than one.
Similarly, we find that the systematic uncertainties resulting from different fitting ranges
are subdominant to the statistical ones and thus are not included in the final results.
The $\eta'$ mass we obtain is $m_{\eta'}=0.992(76)$ GeV on the 64I lattice and $m_{\eta'}=0.926(60)$ GeV on the 48I lattice, respectively.
Following the same argument and strategy as in the $\eta$ case, our final prediction is $m_{\eta'}=0.952(47)(40)$ GeV with the second error being the systematic uncertainty of lattice discretization effects, which is reasonably consistent with the physical value 957.78(6) MeV~\cite{ParticleDataGroup:2024cfk}.

\subsection{mixing angle} 
The mixing of $\rm SU(3)$ singlet state $\eta_1$ and the neutral octet state $\eta_8$ due to the breaking of the flavor symmetry results in the physical states $\eta$ and $\eta'$.
The corresponding mixing angles measure such mixing. 
In the leading order, a single angle describes the mixing. 
Modern treatment of the mixing is based on the parametrization of the matrix elements with two mixing angles as~\cite{Gan:2020aco} 
\begin{equation}
    \left\{
    \begin{array}{cc}
        F^1_{\eta'} = F^1 \cos\theta_1, & F^8_{\eta'} = F^8 \sin\theta_8\\
        F^1_\eta = -F^1 \sin\theta_1, & F^8_\eta = F^8 \cos\theta_8
    \end{array},
    \right.
\end{equation}
where $F^8_\eta = \langle 0 |O_8| \eta\rangle$, 
$F^8_{\eta'} = \langle 0 |O_8| \eta'\rangle$, 
$F^1_\eta = \langle 0 |O_1| \eta\rangle$, and
$F^1_{\eta'} = \langle 0 |O_1| \eta'\rangle$ with
$O_8$ and $O_1$ being the flavor octet and singlet axial-vector currents, respectively.
In our case, however, we use $O_1=q$, 
and thus we arrive at
\begin{equation}
    \tan \theta_1=-\frac{\langle 0|q|\eta\rangle}{\langle 0|q|\eta'\rangle}.
    \label{theta1}
\end{equation} 
In the two-state fits, the spectral weight factors are related to the matrix elements as
\begin{equation}
    A_{\eta/\eta'} =\left|\left\langle 0\left|q\right| \eta/\eta'\right\rangle\right|^{2}.
\end{equation}
Therefore, we have the mixing angle defined using the topological charge operator as
\begin{equation}\label{eq:theta1}
    \theta_1=-\arctan{\sqrt{{A_{\eta}}/{A_{\eta'}}}}.
\end{equation}
\begin{figure}[htbp]
    \centering \includegraphics[width=0.35\textwidth]{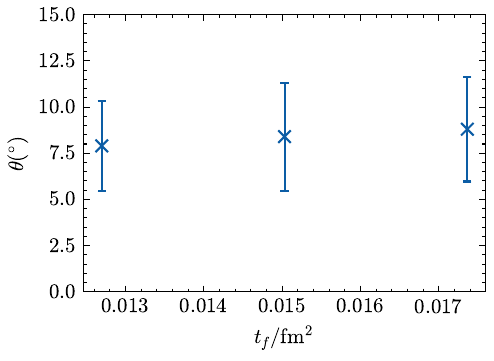}
    \caption{The mixing angle $\theta_1$ as a function of different flow time.
    The error bars show the total error including both the systematic and statistical uncertainties.}
    \label{theta_64}
\end{figure}

The values of ${A_{\eta}}$ and $A_{\eta'}$ are from the previously discussed $\eta'$ fittings.
We determine $\theta_1=-10.7(4.1)^\circ$ on the 64I lattice and $\theta_1=-8.3(2.4)^\circ$ on the 48I lattice.
Still, we use a constant fit for the final prediction and take the difference between the constant-fit result and the value obtained on the finer lattice as the systematic uncertainty of lattice discretization. 
Our final prediction reads $\theta_1 = -8.9(2.1)(1.8)^\circ$ with the second error being the total systematic uncertainty.
Since this study uses topological charge operators, only $\theta_1$ can be determined.

To check the possible flow time dependence of the matrix elements, we plot the numerical results (final results with total error including both the systematic and statistical uncertainties) of the mixing angle $\theta_1$ as a function of the flow time in Fig.~\ref{theta_64}.
As mentioned before, the topological charge operators are properly defined only with large enough Wilson flow times, and the results are indeed quite stable w.r.t. flow times when $t_f>(0.1~{\rm fm})^2$. We find no significant dependence of our results on the flow time within the current precision.

\section{summary and discussion}
\label{sec4}

This work, for the first time, extracts both the $\eta$ and $\eta'$ masses and also the mixing angle $\theta_1$ by using topological charge operators. 
We obtain $m_\eta = 0.505(72)(75)$ GeV, $m_{\eta'}=0.952(47)(40)$ GeV, and $\theta_1 = -8.9(2.1)(1.8)^\circ$, where the first error is the statistical error and the second one is the systematic error of lattice discretization effects.

\begin{figure}[htbp]
    \vspace{0.2cm}
    \centering \includegraphics[width=0.48\textwidth]{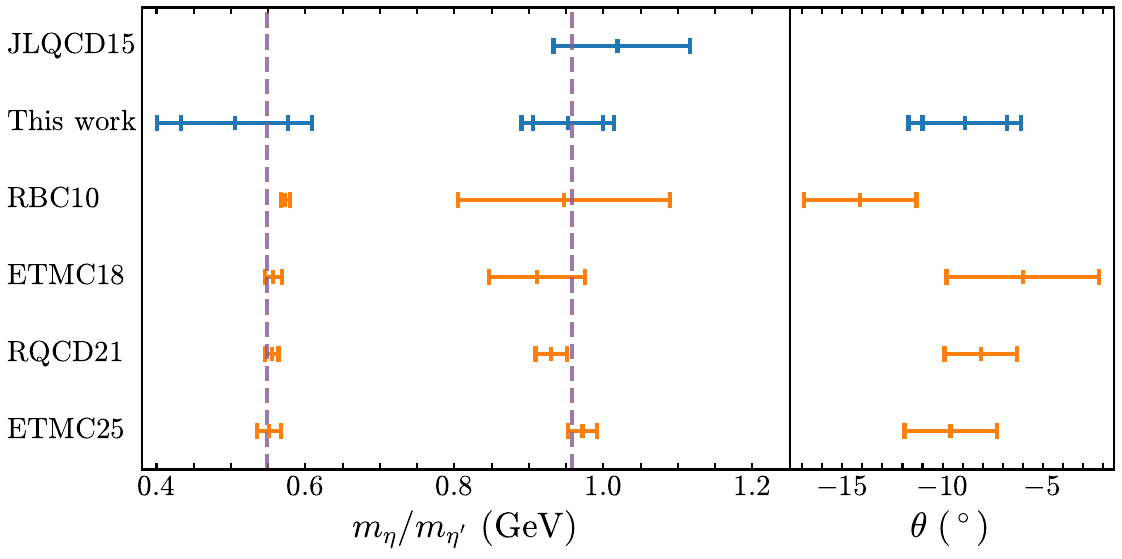}
    \caption{Comparison of the $\eta$ and $\eta'$ masses and the mixing angles from different lattice calculations. 
    The purple vertical dashed lines represent the PDG~\cite{ParticleDataGroup:2024cfk} values of $m_{\eta}$ and $m_{\eta'}$. 
    We collect the results from
    JLQCD15~\cite{Fukaya:2015ara}, 
    RBC10~\cite{Christ:2010dd},
    ETMC18~\cite{Ottnad:2017bjt},
    RQCD21~\cite{Bali:2021qem},  
    and ETMC25~\cite{Ottnad:2025zxq}, which used $N_f$ = 2+1 or $N_f$ = 2+1+1-flavor gauge ensembles and reported both the masses and mixing angles (except for the JLQCD15 study).
    The blue and orange colors represent the results using the topological charge operator and the quark bilinear operators, respectively.
    Some works only reported the mixing angle $\phi$ in the light-strange flavor basis, and the values of $\theta_1$ quoted here are converted.}
    \label{fig:compare_results}
\end{figure}

A comparison with other lattice results is summarized in Fig.~\ref{fig:compare_results} regarding the $\eta$ and $\eta'$ masses and the mixing angle $\theta_1$.
Since some works only reported the mixing angle $\phi$ in the light-strange flavor basis, we convert $\theta_1$ as $\phi$ to $\theta_1 = \phi - \arctan\left(\frac{\sqrt{2}F_l}{F_s}\right)$, where $F_l$ and $F_s$ are related to the decay constants of $\eta$ and $\eta'$ with light and strange quark axial vector currents, respectively.
In Fig.~\ref{fig:compare_results}, the blue and orange points represent the results using the topological charge operators and the quark bilinear operators, respectively.
Compared with the studies using quark operators, 
we get all consistent results including the mixing angle.
The error of the $\eta$ mass is larger, but the $\eta'$ mass 
and the mixing angle defined by the topological charge operators has relatively high precision. 
This can be understood by noticing the fact that the topological charge operator is a flavor-singlet operator which couples primarily to the $\eta'$ state ($\left|{A_\eta}/{A_{\eta'}}\right|=\tan^2\theta_1\sim0.035$), and one needs to go to very large $\hat{r}$ to reach the ground state $\eta$. 
However, since using topological charge operators avoids the calculations of quark propagators, it offers special advantages for studying the $\eta$ and $\eta'$ mesons. 
Future calculations using more gauge configurations are promising to achieve greater precision than using quark operators.

\section*{Acknowledgements}
We thank the RBC/UKQCD collaboration for sharing their DWF gauge configurations. 
We thank Prof.\ Ying Chen and Xu Feng for helpful discussions.
The numerical calculations are performed using the GWU code~\cite{Alexandru:2011ee, Alexandru:2011sc} on the Southern Nuclear Science Computing Center (SNSC) and the Xiangjiang-1 cluster at Hunan Normal University (Changsha).
This work is supported by the National Natural Science Foundation of China (NSFC) under Grant Nos.\ 12175073, 12222503, 12175063, 12105108, and 12205106.
J. L. also acknowledges the support of the Natural Science Foundation of Basic and Applied Basic Research of Guangdong Province under Grant No.\ 2023A1515012712.
L. C. Gui is also supported by the Hunan Provincial Natural Science Foundation No.\ 2023JJ30380 and the Hunan Provincial Department of Education under Grant No.\ 20A310.

\bibliography{bib.bib}
\bibliographystyle{apsrev4-2}

\end{document}